\documentclass[aps,prl,twocolumn,shortbibliography]{revtex4-2}
\usepackage{amsmath}
\usepackage{upgreek}
\usepackage{longtable}
\usepackage{amssymb}
\usepackage{graphicx}
\usepackage{mathrsfs}
\usepackage{ntheorem}
\usepackage{mathtools}
\usepackage{enumerate}
\usepackage{enumitem}
\usepackage[T1]{fontenc}
\usepackage{physics}
\usepackage{overpic}
\usepackage{epstopdf}
\usepackage{array}
 \usepackage{multirow}
 \usepackage{array}

\usepackage{bm}
\usepackage{color,soul}
\usepackage[bookmarks=false]{hyperref}
\usepackage{xcolor}
\hypersetup{colorlinks=true,citecolor=blue,
linkcolor=blue,urlcolor=blue,pdfstartview=FitH,
bookmarksopen=true}
\usepackage{booktabs}
\newcommand{\ot}{\otimes}

\newcommand{\bkt}[2]{\langle{#1}|{#2}\rangle}

\begin{document}
\preprint{APS/123-QED}

\title{Enhancing interferometry using weak value amplification with real weak values}

\author{Jing-Hui Huang$^{1,2}$}
\author{Kyle M. Jordan$^{2}$}
\author{Adetunmise C. Dada$^{3}$}
\author{Xiang-Yun Hu$^{1}$}  \email{Corresponding author :xyhu@cug.edu.cn}
\author{Jeff. S. Lundeen$^{2}$} \email{Corresponding author: jlundeen@uOttawa.ca}
\address{$^{1}$ Hubei Subsurface Multi-scale Imaging Key Laboratory, School of Geophysics and Geomatics, China University of Geosciences, Wuhan 430074, China }
\address{$^{2}$Department of Physics and {Nexus for Quantum Technologies}, University of Ottawa, 25 Templeton Street, Ottawa, Ontario, Canada K1N 6N5 }
\address{$^{3}$School of Physics and Astronomy, University of Glasgow, Glasgow G12 8QQ, UK }

\date{\textbf{Accepted (5 February 2025) in Physical Review Letters}}

\begin{abstract}
We introduce an ultra-sensitive interferometric protocol that combines weak value amplification (WVA) with traditional interferometry. This protocol WVA + interferometry uses weak value amplification of the relative delay between two paths to enhance interferometric sensitivity. As an example, we demonstrate a proof-of-principle experiment that achieves few-attosecond timing resolution (few nanometer path length resolution) with a double-slit interferometer using only common optical components. Since our example uses only the spatial shift of double-slit interference fringes, its precision is not limited by the timing resolution of the detectors, {but is instead limited by the fundamental shot noise associated with classical light and the diminished technical noise}. We experimentally demonstrate that the signal-to-noise ratio can be improved by one to two orders of magnitude relative to a measurement that does not use WVA. Two key conclusions are drawn: (i) Most conventional interferometric techniques primarily rely on determining the path difference (time delay or longitudinal phase), with their precision constrained by technical noise. Our protocol offers a robust solution for minimizing the technical noise in traditional interferometry, with precision in principle approaching the shot-noise limit. (ii) Although WVA has achieved significant advancements in ultra-sensitive longitudinal phase measurement, its applicability is constrained by the need for broad spectral bandwidths and high-resolution spectrometers. Contrary to previous assumptions, we demonstrate that quantum-limited WVA time delay measurements are achievable with narrowband light using real weak values. Thus, the cost-effectiveness and practicality of the proposed WVA + interferometry protocol using narrowband light broaden the scope of WVA applications. This protocol holds potential for broad applications in optical metrology, quantum optics and quantum information, biomedical imaging, and interferometric telescopes for astrophysics.
\end{abstract}


\maketitle
{{$Introduction.$}\,\,}Interferometry is widely used for making ultra-sensitive measurements of time delays. Examples include the Sagnac interferometer~\cite{PhysRevLett.104.251102} for rotation-induced delays, the Mach-Zehnder interferometer~\cite{Sorianello2018} for detecting refractive index changes in a sample, and Michelson interferometers for gravitational wave detection~\cite{PhysRevLett.116.061102}.  Detecting few-nanometer path changes, which corresponds to attosecond time delay measurements, is a crucial tool in cellular biology and studying monatomic layer two-dimensional materials~\cite{Chen:21}. 
Accurate time delay estimation is also crucial in pump-probe~\cite{PhysRevLett.102.213904} and femtosecond interferometry~\cite{doi:10.1126/science.1090277}.
There are two contributions that limit the precision in interferometers, technical noise and fundamental quantum noise.
One outcome of quantum metrology has been to design interferometers that reduce the latter contribution, for example, by using entangled photons~\cite{RevModPhys.81.865,PhysRevLett.113.030401} or squeezed light~\cite{PhysRevLett.104.251102}. 
However, technical noise is actually the primary limiting factor in most interferometers, including scientific ones, not to mention commercial interferometric sensors.

{
Weak value amplification (WVA)~\cite{RevModPhys.86.307,PhysRevLett.60.1351,PhysRevLett.66.1107,PhysRevLett.132.250802,XU2024100518} has gained substantial attention as a potent technique for amplifying small physical effects, such as 
photonic spin hall effect~\cite{doi:10.1126/science.1152697,PhysRevA.84.043806,PhysRevApplied.13.014057}, 
transverse optical deflections~\cite{PhysRevLett.102.173601}, 
charge sensing~\cite{PhysRevLett.106.080405},
single-photon nonlinearity~\cite{PhysRevLett.107.133603,Hallaji2017},
angular rotations~\cite{PhysRevLett.112.200401}, and longitudinal phase shifts~\cite{PhysRevLett.105.010405,PhysRevLett.110.083605,PhysRevLett.111.033604}. Meanwhile, numerous WVA-based metrology protocols~\cite{PhysRevA.94.053843,Turek_2015,PhysRevLett.117.230801,PhysRevLett.116.180401,Chen2018,PhysRevA.99.013801,Yin2021,PhysRevLett.126.220801,PhysRevA.105.043508,PhysRevA.106.022619,PhysRevA.107.042601,PhysRevA.108.032608}, especially the cyclic WVA protocols~\cite{PhysRevLett.117.230801,PhysRevLett.126.220801,PhysRevA.108.032608} show significant improvement in the signal-to-noise ratio (SNR).
}
{By employing the Fisher information analysis~\cite{PhysRevLett.115.120401,PhysRevX.4.011032,PhysRevX.4.011031,PhysRevA.106.022619,PhysRevA.107.042601,PhysRevA.105.013718}, prior experiments demonstrated that WVA is capable of achieving the shot-noise limit~\cite{PhysRevLett.125.080501}.}

Here, we propose a new interferometric protocol, WVA+interferometry, that enables ultra-sensitive longitudinal shift measurements {and diminishes technical noise.} These shifts can be thought of interchangeably as time delays or path-length changes. In standard interferometry, when using light with a central optical frequency of $f$, a delay $\tau$ shifts the interferometer output intensity by a fraction $f\tau$ of an interference fringe, moving from, for example, destructive interference towards constructive interference.
Thus, the fringe shift is used to estimate the delay $\tau$. Noise, however, inhibits the ability to accurately estimate the interferometric output intensity and, as a result, $\tau$. 
Shot noise sets the fundamental quantum limit for estimation of delays where a classical-like state (e.g. coherent state) is input to the interferometer. 
{Furthermore, experiments with single photons observe the same interference patterns as classical light~\cite{RevModPhys.92.035005}, and experience statistical noise of the same magnitude as shot-noise~\cite{PhysRevLett.96.010401}, so that our results are relevant also in the single photon regime.}
{While most WVA protocols implicitly rely on interference effects, applications of WVA have so far mainly focused on amplifying a parameter that is then directly measured. Meanwhile, precise optical measurements typically rely on traditional interferometry~\cite{PhysRevLett.104.251102,Sorianello2018,PhysRevLett.116.061102}.
It is therefore natural to ask whether WVA can be used to further enhance the precision of a traditional interferometer. Our protocol demonstrates that this is indeed possible: by amplifying the phase shift inside an interferometer, we marry the precision of traditional interferometry with the noise-reduction advantages of WVA~\cite{PhysRevLett.115.120401,PhysRevX.4.011031,PhysRevA.94.012329,PhysRevA.107.052214,PhysRevA.97.033851,Xia2023}. 
Furthermore, it enables WVA-enhanced measurement of delay and phase using a narrowband probe and conventional optics, which was previously thought to be impossible~\cite{PhysRevLett.105.010405}}.

{Fundamentally, optical measurements of a phase $\phi_0$ are limited by the total number of input photons $N_\text{in}$ according to $\Delta\phi_0 \geq 1/\sqrt{N_\text{in}}$, which is an expression of shot noise. 
However, in applications that are limited by technical noise, this limit is rarely reached. While one can in principle remove most technical noise by suitable engineering, as was done at LIGO~\cite{PhysRevLett.129.061104,PhysRevX.13.041039}, it is often more practical to instead seek techniques that can suppress the harmful effects of existing noise on the final measurement sensitivity. 
Previous work by A. N. Jordan \textit{et al.} has shown that WVA using real weak values can, with suitable post-selection, 
put all of the Fisher information into the post-selected events~\cite{PhysRevX.4.011031}; because of this, the decreased intensity due to post-selection does not inherently limit the precision of WVA. Furthermore, the enhanced SNR can reduce the effects of technical noise, so that the primary application of WVA is as a simple method to improve the SNR of a non-optimal sensor~\cite{PhysRevLett.125.080501,PhysRevLett.128.040503}.}
Our protocol combines WVA with interferometry by amplifying the delay between the two interferometer paths to be $A_{w}\tau$, where the amplification factor $A_{w}$ is known as the weak value. This results in a proportionately amplified fringe shift $fA_{w}\tau$, which is then used to estimate $\tau$. We present a proof-of-principle experiment, using a double-slit interferometer as an example. The delay amplification results in an amplified spatial shift of the double-slit interference pattern, enabling few-attosecond time delay measurement with enhanced precision when compared to conventional techniques.

The optimal way to use WVA has been a matter of debate, partly because the weak value $A_{w}$ is a complex number. For a temporal delay $\tau$, the real part of $A_{w}$ amplifies the temporal shift of an optical pulse, whereas the imaginary part shifts the center frequency of the pulse by an amount proportional to $\tau$. One can design the WVA so that $A_{w}$ is either purely real or purely imaginary. An early work by Brunner \textit{et al}.~\cite{PhysRevLett.105.010405} stated that only a purely imaginary weak value leads to any advantage over standard techniques such as interferometry.
Since the signal manifests in a spectral shift, neither the pulse length nor the detector timing resolution limits measurement of $A_{w}\tau$. In comparison, the temporal shift induced by a real weak value is typically much shorter than the timing resolution of realistic photodetectors, and so could not be used for enhanced sensitivity to time delays. Xu \textit{et al}.~\cite{PhysRevLett.111.033604} then reported a high-precision delay measurement using a purely imaginary weak value and a white light source.
Hence, when measuring longitudinal shifts with WVA, the predominant choice has been to use an imaginary weak value. 
On the other hand, real weak values are often used to amplify transverse spatial shifts~\cite{PhysRevLett.118.070802} and can achieve quantum-limited precision with imperfect, potentially noisy, detectors~\cite{PhysRevLett.125.080501}.
Our example protocol combines the spatial interferometry of the double-slit with amplification using real weak values. We will show that the combination eliminates the need for either a broadband or pulsed source of light (and respective high-resolution spectrometer or high-speed photodetector) while retaining the noise advantages previously demonstrated with spatial WVA.

\begin{figure*}[t]
\centering
\includegraphics[width=0.98\textwidth]{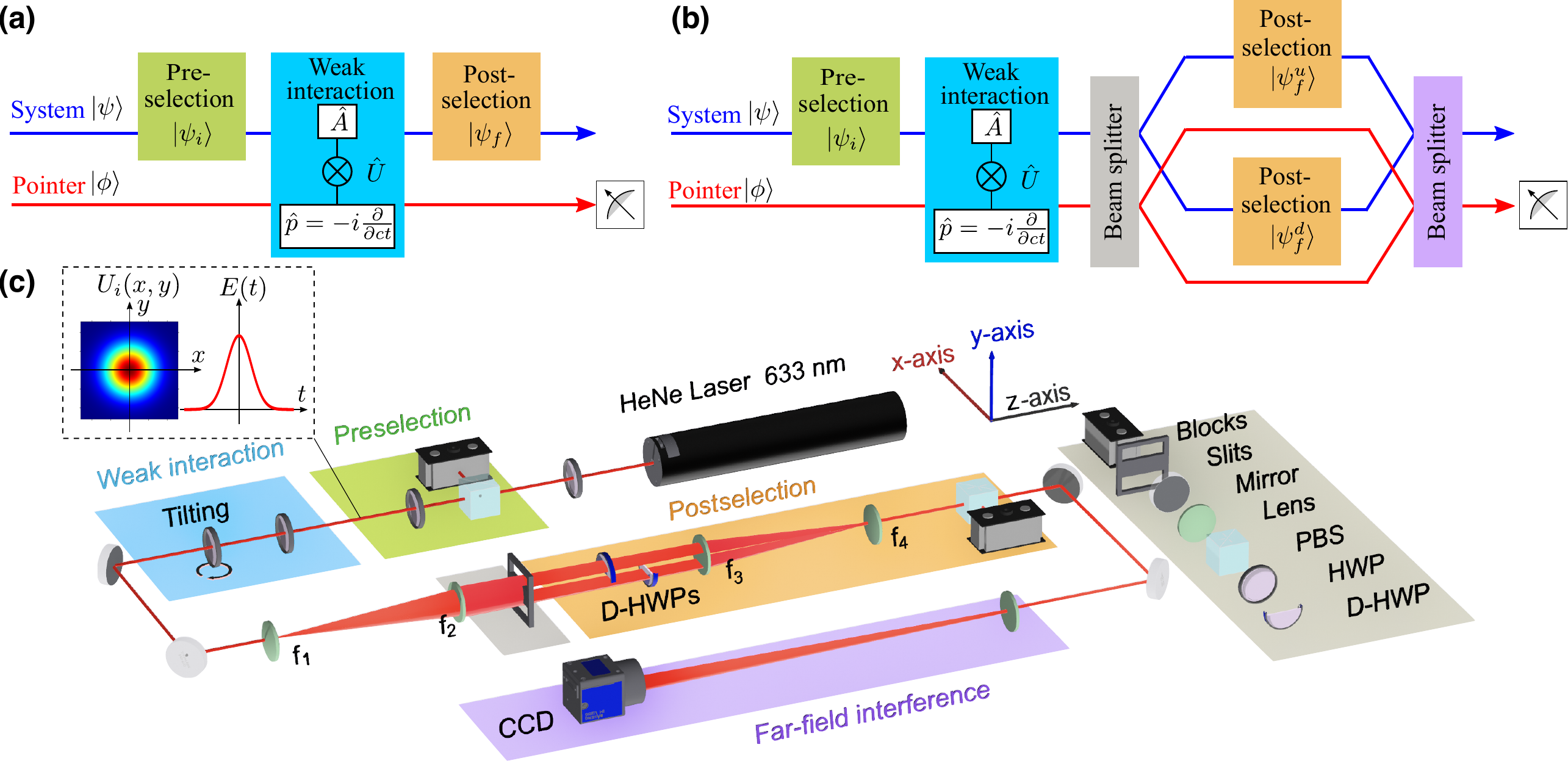}
\caption{\label{Fig:Schemes_model}
Study of the proposed WVA+interferometry protocol.
Comparison of (a) the standard WVA with (b) the proposed WVA+interferometry.
The weak interaction between the system $ \ket{\psi_{}}$ and the pointer $ \ket{\phi_{}}$ is described as the unitary operator ${\hat{U}} = {\rm exp}(-i g \hat{A} \ot \hat{p})$.
(c) Experimental setup.
{
The two pointers undergo identical pre-selection and weak interaction but two different post-selections. By setting the post-selected state in the upper arm $u$ to be different from the post-selected state in the lower arm $d$, each arm experiences a different amplification $A_{w}^{u,d}$ thereby increasing the interferometer delay.}
}
\end{figure*}

{{$Theoretical \, \, framework.$}\,\,}
Fig.~\ref{Fig:Schemes_model}(a) depicts the scheme for standard WVA. In it, an interaction described by the unitary evolution $\hat{U}=e^{-ig\hat{A}\ot\hat{p}}$ couples a two-level “system” $\ket{\psi}$ to a “pointer” $\ket{\phi}$, where $\hat{A}$ is the observable quantity and $g$, the coupling strength, is the parameter we wish to estimate. The momentum operator $\hat{p}$ acts on the pointer state $\ket{\phi}$ in the coordinate representation with its position $q$. Thus, for a particular value of $A=a$, ${\hat{U}}$ acts as a translation operator and shifts the pointer by $\Delta q=ag$. The goal of WVA is to amplify this shift.
In the weak limit $\Delta q\ll1$, if the system begins in polarization state $\ket{\psi_{i}}$ (“pre-selection”) and undergoes ${\hat{U}}$, then the light that passes through a polarizer projecting onto $\ket{\psi_{f}}$ (“post-selection”) will have an average pointer shift of $\Delta q=gA_{w}$, with $A_{w}=\bra{\psi_{f}}\hat{A}\ket{\psi_{i}}/\bkt{\psi_{f}}{\psi_{i}}$. If the pre- and post-selected states are almost orthogonal, the denominator is small, resulting in $A_{w}\gg1$, i.e., leading to amplification of the weak value.

Weak value amplification can amplify the delay inside an interferometer a few different ways. In the conceptually simplest way, the full WVA scheme is inside a single interferometer path. Alternately, solely the pre- or post-selection occurs inside the interferometer. We depict the latter in Fig.~\ref{Fig:Schemes_model}(b). 
The initial system is prepared as $\ket{\psi_{i}}=\sin{(\pi/4)}\ket{H} + \cos{(\pi/4)}\ket{V}$, where $H$ and $V$ are horizontal and vertical polarizations. 
We aim to measure the birefringent time delay $\tau$ introduced between the two polarization states. The two post-selections project onto states: $| {\psi_{f}^{u,d}} \rangle =  e^{-i {\omega} \tau /2}\sin{( 3\pi/4 +\beta^{u,d} )} \ket{H} 
+ e^{+i {\omega} \tau /2} \cos{( 3\pi/4 +\beta^{u,d}  )} \ket{V},$
where ${\omega}=2\pi c/\lambda$ denotes the angular frequency and $\lambda$ represents the wavelength of the photon. Throughout the calculations, we use superscripts $u$ and $d$ to denote quantities related to the upper and lower arms, respectively.
Choosing the post-selection angles to be $\beta^{u}$ and $\beta^{d}$ leads to two different weak values:
\begin{equation}
\label{Eq:definition_weakValue}
A_{w}^{u,d}=\frac{\langle \psi^{u,d}_{{f}}| \hat{A}\ket{\psi_{i}}}{\bkt{\psi^{u,d}_{{f}}}{\psi_{i}}}=
\frac{e^{i {\omega} \tau} -  \cot{(3\pi/4 + \beta^{u,d})} }{ e^{i {\omega} \tau} +  \cot{( 3\pi/4 + \beta^{u,d}} ) }.
\end{equation}
The post-selections result in amplified shifts $\delta t^{u,d}={\mathcal{R}e}[A_{w}^{u,d}]\tau$. Choosing $\beta^{u,d}$ with opposite signs can obtain the maximum relative shift $|\delta t^{u}-\delta t^{d}|$ for a given magnitude of $\beta$.
Setting $\beta^{u,d}=\pm45^{\circ}$ results in $A_{w}^{u,d}\approx\pm1$, representing a standard (no-WVA) interferometer. 
\begin{figure*}[t]
\centering
\includegraphics[width=0.98\textwidth]{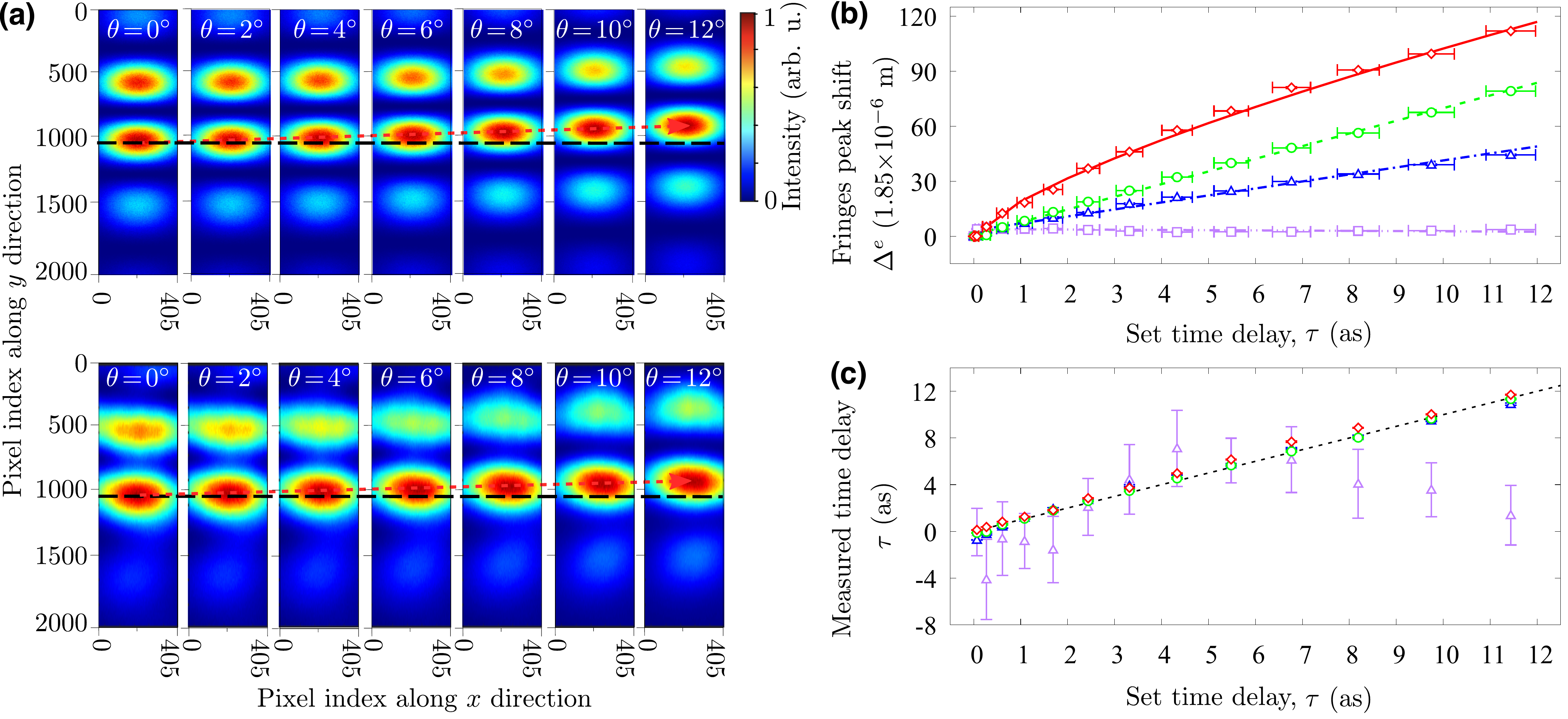}
\caption{\label{Fig:CCDiamges_aand_PeakShifts}
Simulation and experimental results. (a) Simulation (upper) and experimental (lower) CCD images vary with the time-delay waveplate tilting angle $\theta$ when setting $\beta_{u,d}=\pm1.6^{\circ}$. The red arrow line is a guide for shifts. 
{
The simulation, which is further explained in SM, incorporates a non-zero initial time delay to account for any extra path length caused by misalignment of the surfaces of the two D-HWPs. 
}
(b) Experimental fringe shifts $\Delta^{e}$ with varying time delay $\tau$ and various post-selection angle $\beta^{u,d}$.
{Lines and points with error bars represent the calibration curves and experimental results, respectively.}
{The horizontal error bars correspond to the uncertainty of $\delta \theta=\pm 0.15^{\circ}$ in the set time delay; 
the vertical error bars correspond to the uncertainty in estimating the fringe shifts.}
(c) {Measured} time delay
vs. set time delay. The red, green, blue and purple data represent $\beta^{u,d}=\pm1.6^{\circ}$, $\beta^{u,d}=\pm3.3^{\circ}$, $\beta^{u,d}=\pm6.6^{\circ}$ and $\beta^{u,d}=\pm45^{\circ}$, respectively. 
}
\end{figure*}

For photons propagating along $z$-coordinate with three-dimensional coordinates $\Vec{q}=(x,y,z=ct)$, a representation of the pointer state is $\langle \Vec{q} \ket{\phi}  = U(x, y) {E(t)}$,
where ${E(t)}$ is the temporal envelope and $U(x,y)$ represents the transversal profile. Here, $c$ denotes the speed of light and the $t$ denotes the time offset for photons reaching a detector.
Initially, the pointer state $\langle{\Vec{q}}\ket{\phi_{i}}=U_{i}(x,y){E(t)}$ generated by a laser is assumed to have a Gaussian profile characterized by $U_{i}(x,y)={\rm exp}[-(x^{2}+y^{2})/{\sigma_{xy}}^{2}]$ with a beam's width of $\sigma_{xy}$. {We model our monochromatic light as a train of overlapping pulses with a duration equal to the coherence time of the laser, so that a single pulse has a field} ${E(t)}=E_{0}\, \exp[-(t^{2})/\sigma_{t}^{2}]$. The constant $E_{0}$ is the strength of the electric field, and $\sigma_{t}$ characterizes the pulse length. The collimated beam is split into two pointers $\langle{\Vec{q}}|{\phi_{i}^{u,d}}\rangle=E^{u,d}(t)U^{u,d}(x_{1},y_{1})$. These two beams pass through a lens with focal length $f_{d}$, resulting in complex amplitudes $U^{u,d}(x_{2},y_{2})$ at the plane $(x_{2},y_{2})$. The final state $|{\phi_{f}^{t}}\rangle$ describing the interference of the two pointers is calculated by:
\begin{eqnarray}
\label{Eq:synthesized-electric-field}
\begin{split}
\langle {\Vec{q}} \ket{\phi_{f}^{t}}
 = &   \frac{E^{u}(t-  {\mathcal Re}[A^{u}_{w}] \tau)}{|A^{u}_{w}|}  U^{u}(x_{2},y_{2}) e^{i {\omega} {\mathcal Re}[A^{u}_{w}] \tau} \\
 &+    \frac{E^{d}(t-  {\mathcal Re}[A^{d}_{w}] \tau)}{|A^{d}_{w}|}  U^{d}(x_{2},y_{2}) e^{i {\omega} {\mathcal Re}[A^{d}_{w}] \tau}.
\end{split}
\end{eqnarray}
The time delay $\tau$ can be estimated from the fringe distribution $I_{CCD}(x,y)\propto|\langle{\Vec{q}}|{\phi_{f}^{t}\rangle}|^{2}$~\cite{Equation-final-interference}.
In addition, Eq.~(\ref{Eq:synthesized-electric-field}) is applicable to arbitrary envelopes $E(t)$ as long as the parameter $\tau$ lies within the weak interaction regime $(\tau\ll\sigma_{t})$.
Here, we apply a CW laser to estimate the relative time delay between two paths, which is due to the birefringent delay of a crystal~\cite{PhysRevLett.111.033604}. 

{{$Experiment.$}\,\,}
The experimental setup is shown in Fig.~\ref{Fig:Schemes_model}(c). We employ a HeNe Laser with a 0.65 mm beam diameter. Laser power is controlled using a half-wave plate (HWP). The photons then pass through a polarizing beam splitter (PBS) and a HWP, preparing the pre-selected state.
We use two crossed true zero-order HWPs to introduce a controllable birefringent time delay~\cite{PhysRevLett.111.033604}.
The optical axes of the first and second HWPs are respectively along the $x$-axis and $y$-axis.
By tilting the second HWP by an angle $\theta$ around the $y$-axis, we introduce a delay between the $x$ and $y$ polarizations of $\tau \approx {\pi \theta ^{2}}/{2 n_{0}^{2} {\omega}}$,
where $n_{0}=1.54$ represents the refractive index of the HWPs~\cite{PhysRevLett.111.033604}. To enable spatial splitting, a combination of a beam expander $(f_{1} = 25.5\, \mathrm{mm}, f_{2} = 300\, \mathrm{mm})$ and a reversed beam expander $(f_{3} = 300\, \mathrm{mm}, f_{4} = 25.5\, \mathrm{mm})$ is used.
Within this region, we introduce a two slits with a width of 50 mm and a gap of $D_{1} = 5\, \mathrm{mm}$.
Two D-shaped HWPs followed by a subsequent PBS carry out the two post-selections.
Far-field interference is generated by a lens ($f_{d} = 1\, \mathrm{m}$) and is detected by a CCD~\cite{Ref_CCD}.

In this Letter, the fringe shifts $\Delta^{e,s}$ are calculated from the per-pixel CCD counts $k_{mn}^{e,s}$ with a distribution $p(k_{mn}^{e,s}|\tau,X)$, where $X$ denotes all available information about the CCD. The superscripts “$s$” and “$e$” are used to distinguish simulation and experiment, respectively. 
We use an image registration algorithm with upsampled cross-correlation \cite{Guizar-Sicairos:08} to calculate $\Delta^{e,s}(\tau)$ along the $y$ direction.
Then, the classical Fisher information (CFI) for estimating $\tau$ is calculated as:
 \begin{equation} \begin{split}
\label{Eq:define_FisherInformation}
F^{e,s}
 = \sum\limits_{m} p(K^{e,s}_{m} | \tau, X)
 \times \left[  \frac{\partial}{\partial \tau} {{\rm ln} {\,} p(K^{e,s}_{m} | \tau, X)}  \right]^{2},
\end{split}
\end{equation}
with $K_{m}^{e,s}=\sum_{n}k_{mn}^{e,s}$, where $m$ and $n$ are the pixel indices along $y$- and $x$- directions, respectively. 
The reason why we sum over pixels is that tilting the time-delay waveplate around $y$- axis inevitably leads to horizontal shifts due to refraction at the HWP [see Fig.~\ref{Fig:CCDiamges_aand_PeakShifts}(a)].
Calculations of the theoretical CFI assume that only shot noise is present in Supplementary Material (SM)~\cite{Link_Supplemental_Material}, and so serve as an upper bound on the precision of realistic measurements including technical noise~\cite{10.1093/acprof:oso/9780198563617.001.0001}. 

We present the experimentally observed fringe shift $\Delta^{e}(\tau)$ alongside the theoretical $\Delta^{s}(\tau)$ in Fig.~1(b) of SM.
{At larger post-selection angles $|\beta^{u,d}| > 1.6^{\circ}$, minor deviations between experimental data and theoretical predictions are observed. The discrepancy arises primarily due to manual adjustments of the waveplate angle $\theta$, contributing to horizontal error bars in Fig.~\ref{Fig:CCDiamges_aand_PeakShifts}(b)~\cite{Erorr_set_time_delay}, as well as slight variations in the post-selection angle. 
At the smallest $\beta^{u,d}=\pm1.6^{\circ}$, the experimental data are generally lower than the theory but the fringe shifts still show amplification. This mismatch is due to an increased sensitivity to experimental imperfections, and is common when conducting WVA at nearly orthogonal pre- and post-selected states (see Ref.~\cite{PhysRevLett.102.173601}).
To eliminate the discrepancy, we fit each experimental data set in Fig.~\ref{Fig:CCDiamges_aand_PeakShifts}(b) to a calibration curve~\cite{Data-fitting-method}, which then functions as our calibration of time delay per microns of fringe shift for each amplification factor (i.e., orthogonality $\beta^{u,d}$). 
The calibration curve depicted in Fig.~\ref{Fig:CCDiamges_aand_PeakShifts}(b) is then used to perform the time delay estimation in Fig.~\ref{Fig:CCDiamges_aand_PeakShifts}(c).
}

\begin{figure}[t]
\centerline{\includegraphics[scale=0.26,angle=0]{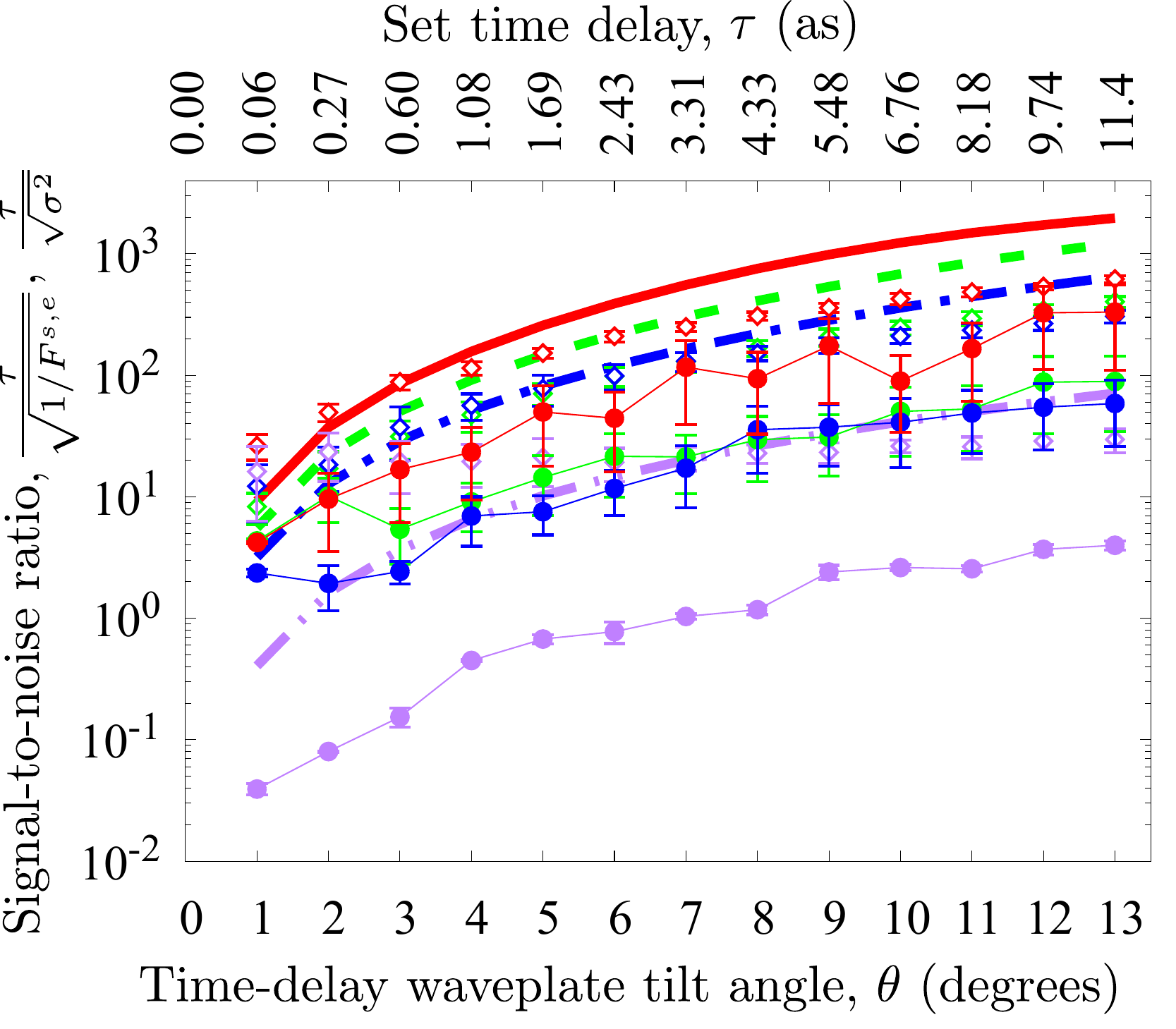}}
\caption{\label{Fig:Result_only_CFI}
Comparison of signal-to-noise ratio to the shot noise limit. Theoretical shot noise limit ${\tau}{\sqrt{F^{s}}}$ (thicker lines), experimental shot noise limit ${\tau}{\sqrt{F^{e}}}$ (points with error bars) and the experimental SNR ${\tau/{\sigma}}$ ({guide lines and points with  statistical error bars}). The red, green, blue and purple data represent $\beta^{u,d}=\pm1.6^{\circ}$, $\beta^{u,d}=\pm3.3^{\circ}$, $\beta^{u,d}=\pm6.6^{\circ}$ and $\beta^{u,d}=\pm45^{\circ}$, respectively.
}
\end{figure}

{
Figure~\ref{Fig:CCDiamges_aand_PeakShifts}(c) demonstrates that our protocol is capable of considerably mitigating technical noise. The estimation results with $\beta^{u,d}=\pm45^{\circ}$ represent the traditional interferometer and exhibit significant oscillation and error bars. When selecting a smaller post-selection angle, the weak value increases and the error bar decreases, thereby enhancing precision due to the noise-reduction advantages of WVA~\cite{PhysRevLett.115.120401,PhysRevX.4.011031,PhysRevA.94.012329,PhysRevA.107.052214,PhysRevA.97.033851,Xia2023}. 
}

In Fig.~\ref{Fig:Result_only_CFI}, we present the precision in the measurements $\Delta^ {}(\tau)$ by calculating $F^{s,e}(\tau)$ and the actual SNR $\tau/{\sigma}$. Here, ${{\sigma}^{2}}$ is the experimental variance in Fig.~\ref{Fig:CCDiamges_aand_PeakShifts}. The precision limit is defined as the Cram{é}r-Rao bound. This bound determines the maximum precision possible in estimating an unknown parameter, as determined by the statistical behavior of the measured data.

{{$Discussion.$}\,\,}
The experimentally determined values ${\tau}{\sqrt{F^{e}}}$ are generally lower than the simulated curves ${\tau}{\sqrt{F^{s}}}$ as shown in Fig.~\ref{Fig:Result_only_CFI}. We attribute this difference to imperfections in the measured diffraction pattern; the experimental probability distribution $p(K_{m}^{e,s}$) presented in  SM is seen to have lower fringe visibility than is expected from the simulations. 
{Fig.~\ref{Fig:Result_only_CFI} also indicates the experimentally SNRs are much lower than experimentally determined values ${\tau}{\sqrt{F^{e}}}$. 
The reason is that our experiments are inevitably affected by the technical noise.
}

In summary, we have proposed and experimentally demonstrated a way to integrate weak value amplification with precision interferometry. 
As an example, we amplified the delay between light passing through the different slits of a double-slit interferometer, which demonstrated a new efficient method for measuring few-attosecond time delays (see Fig.~\ref{Fig:CCDiamges_aand_PeakShifts}) with SNRs improved by up to two orders of magnitude (see Fig.~\ref{Fig:Result_only_CFI} with $\beta^{u,d}=\pm1.6^{\circ}$).
{This technique could be applied to improve the SNR of phase measurements that are restricted in available bandwidth, such as detectors relying on cavity enhancements~\cite{PhysRevLett.116.061102}.}
Our experiment first achieved this level of precision using narrowband light for WVA, enabling a new class of WVA-based measurements of longitudinal phase.

\vspace{10pt}
\begin{acknowledgments}This study was supported by the NSFC (Grants No.~42327803 No.~42488201 and No.~42220104002), the Canada Research Chairs Program, and the Natural Sciences and Engineering Research Council. 
J.H.H. acknowledges support from the CSC. A.C.D. acknowledges support from the EPSRC, Impact Acceleration Account (Grant No.~EP/R511705/1).
The data that support the findings of this article are openly available~\cite{DVN/0WERLH_2025}.
\end{acknowledgments}

\bibliography{apssamp}

\begin{thebibliography}{59}
\bibitem [{Equ()}]{Equation-final-interference}%
  \BibitemOpen
  \href@noop {} {\bibinfo  {journal} {\textup{The use of two different post-selection angles will cause the two pointers to have different polarization states in the far field, reducing the visibility of interference; however, in the large amplification limit ($|\beta^{u,d}|\ll1$) this effect is small, and we do not include it in our calculations}}\ }\BibitemShut {NoStop}%
\bibitem [{Ref()}]{Ref_CCD}%
  \BibitemOpen
\bibfield  {journal} {  }\href@noop {} {\bibinfo  {journal} {\textup{We use a scientific CCD with a pixel size of $\zeta_s \times \zeta_s = 1.85\, \upmu \mathrm{m} \times 1.85\, \upmu \mathrm{m}$, a pixel bit depth of 8 bits and detection efficiency of $\eta=85.6\%$ }}\ }\BibitemShut {NoStop}%
\bibitem [{Ero()}]{Erorr_set_time_delay}%
  \BibitemOpen
  \href@noop {} {\bibinfo  {journal} {\textup{The error in our set time delay arises from the precision limitations of the Manual Rotation Stage used to adjust the waveplate angles. We used the Vernier scale instead of the micrometer since the stage cannot provide 0-13 degree rotation in the experiment. The vernier scale can provide 5 arcmin ($\approx 0.1^{\circ}$) resolution. However, in practical applications, even with careful manual adjustments, the typical uncertainty might range from $0.1^{\circ}$ to $0.3^{\circ}$ when using the Vernier scale to set angles by hand. The primary sources of uncertainty include parallax error, operator dexterity and friction in the mounting mechanism. Therefore, we include an uncertainty of $\pm 0.15^{\circ}$ to set the time-delay waveplate tilt angle across all experimental data. This error results solely from the precision of the optomechanics used, and so does not reflect a fundamental limitation to the optical protocol itself}}\ }\BibitemShut {NoStop}%
\bibitem [{Dat()}]{Data-fitting-method}%
  \BibitemOpen
\bibfield  {journal} {  }\href@noop {} {\bibinfo  {journal} {\textup{In Fig.~\ref{Fig:CCDiamges_aand_PeakShifts}(b), the data with different $\beta^{u,d}$ were calibrated using power-law fitting with the function $\Delta^{e}(\tau)=a {\tau}^b$. Here, $a$ and $b$ represent the fitting parameters}}\ }\BibitemShut {NoStop}%
\end{thebibliography}

\newpage
\onecolumngrid
\appendix

\vspace{20pt}
\begin{center}\textbf{\Large Enhancing interferometry using weak value amplification with real weak values: Supplementary Material}
\end{center}

This document provides supplementary information to “Enhancing interferometry using weak value amplification with real weak values”.
Here, we present the theoretical classical Fisher information calculation.
Then, {we present more details about the simulation and experimental results. Figure 1(a) displays the 
simulation with a non-zero initial time delay and experimental probability distributions $p(K^{e,s}_{m})$ along $m-$direction pixels for different values of $\theta$.} {Figure 1(b) shows the experimentally observed fringe shift $\Delta^{e}(\tau)$ alongside the theoretically simulated fringe shift $\Delta^{s}(\tau)$.
At the end, we present the calculation of the standard deviation of the experimental signal-to-noise ratio.}

\section{Theoretical classical Fisher information calculation}

The theoretical classical Fisher information (CFI) for estimating $\tau$ is calculated as:
 \begin{equation} \begin{split}
\label{Eq:define_FisherInformation}
F^{s}
 = \sum\limits_{m} p(K^{s}_{m} | \tau, X) 
 \times \left[  \frac{\partial}{\partial \tau} {\rm ln}  \, p(K^{s}_{m} | \tau, X)  \right]^{2}. 
\end{split} 
\end{equation}
Where $k_{mn}^{s}$ represents the theoretical CCD outcomes.
Drawing inspiration from investigations in the context of imperfect CCDs~\cite{PhysRevLett.118.070802,PhysRevLett.125.080501}, 
our simulation begins with the calculation of the theoretical distribution {$p(N^{s}_{mn}| \tau)$ }, which represents the expected average number {$N^{s}_{mn}$} of photons received by the CCD.
Taking into account the proportionality between $N^{s}_{mn}$ and $I_{CCD}(m,n)$, the distribution $p(N^{s}_{mn}| \tau) $ equals the distribution $p(I_d(m,n) | \tau) $.
The total average photon number {$\overline{N^{s}}$} of the coherent beam is a fundamental parameter in our simulation, and it is determined by summing all photons {$N^{s}_{mn}$} at each pixel. 
To obtain the simulation input, we maintain {$\overline{N^{e}}$} by measuring the optical average power ${P^{e}}$ before reaching the CCD. To achieve this, we employ the relationship $\overline{N^{e}}={P^{e}} \times T_{CCD}/(E_{p})$, where $T_{CCD}$ represents the exposure time, and each photon at 632.992 nm carries an energy of $E_{p}=3.318 \times 10^{-19} J$. 
The exposure time is set at $T_{CCD}$ = 650 $\mu$s, 300 $\mu$s, and 100 $\mu$s for $\beta_{u,d}= \pm 1.6^{\circ}$, $ \pm 3.3^{\circ}$, and $ \pm 6.6^{\circ}$, respectively.
The optical power detected before the PBS for post-selection is 2.0 mW.
The average power ${P^{e}}$ with different $\beta_{u,d}$ and various $\theta$ are recorded for the simulation input.
We also conduct a baseline measurement with $\beta_{u,d}= \pm 45^{\circ}$, ${P^{e}}=  $ 7.3 ${\rm \mu W}$ and $T_{CCD} $ = 60 $\mu$s, representing the traditional interferometer without WVA enhancement.
For a coherent beam, the exact number $M^{s}_{mn}$ of the photoelectrons detected by the CCD follows a Poisson distribution~\cite{10.1093/acprof:oso/9780198563617.001.0001}
\begin{equation} \begin{split}
\label{Eq:Posson_distribution_simulation}
p(M^{s}_{mn}| \tau,X)
&=\frac{ (\eta N^{s}_{mn})^{M^{s}_{mn}} e^{-\eta N^{s}_{mn} } }{ M^{s}_{mn} ! }.
\end{split} \end{equation}
This Poisson process generates shot noise, which characterizes the fluctuations in the number of photons detected by CCD due to their independent occurrence. 
In simulation, we focus solely on the shot noise, disregarding other forms of electrical noise and avoiding CCD saturation. 
Thus, we mathematically establish a conditional probability distribution $\mathcal{R}(k_{mn}^{s}| M^{s}_{mn})$ = 1, linking $k_{mn}^{s}$ with $M^{s}_{mn}$. 
Finally, the CCD outcomes $k_{mn}^{s}$ closely approximate $M^{s}_{mn}$ and are obtained through the Monte Carlo simulation of the Poisson process.

\section{Simulation and experimental results}

Our experiment show that the experimental CCD image exhibits asymmetry when measuring at $\tau=0\, \mathrm{as}$, which can be attributed to a non-zero initial time delay arising from a slight misalignment of the two D-HWPs' surfaces. 
The slight misalignment of the two D-HWPs’ surfaces can not be easily eliminated experimentally. The true zero-order waveplate we used is made of a quartz plate and a BK7 substrate with a total thickness of 1.1±0.2 mm. When the misalignment angle of the two D-HWPs’ surfaces varies from 0.01 rad to 0.05 rad, the initial time delay $t_0$ changes from 111 as to 2700 as. Therefore, a slight misalignment will lead to a non-ignorable initial time delay.
We further test this assumption by adding a time delay $t_0$ to our simulation. Mathematically, the final state $|{\phi_{f}^{t}}\rangle$ by setting an initial time delay $t_0$ between the two paths is calculated by:
\begin{eqnarray}
\label{Eq:synthesized-electric-field}
\begin{split}
\langle {\Vec{q}} \ket{\phi_{f}^{t}}
 = &   \frac{E^{u}(t-  {\mathcal Re}[A^{u}_{w}] \tau-t_0) }{|A^{u}_{w}|}  U^{u}(x_{2},y_{2}) e^{i {\omega} {\mathcal Re}[A^{u}_{w}] \tau}  e^{i {\omega}  t_0}
 +    \frac{E^{d}(t-  {\mathcal Re}[A^{d}_{w}] \tau)}{|A^{d}_{w}|}  U^{d}(x_{2},y_{2}) e^{i {\omega} {\mathcal Re}[A^{d}_{w}] \tau}.
\end{split}
\end{eqnarray}
{
We plot theoretical modeling in Fig. 2(a) of the main text where the phase offset is correct by setting $t_0$= 900 as, as found from the full set of experimental curves. The agreement, while not perfect, is qualitatively correct, particularly for the fringe shift.
}

\begin{figure*}[t]
\centering
\includegraphics[width=0.95\textwidth]{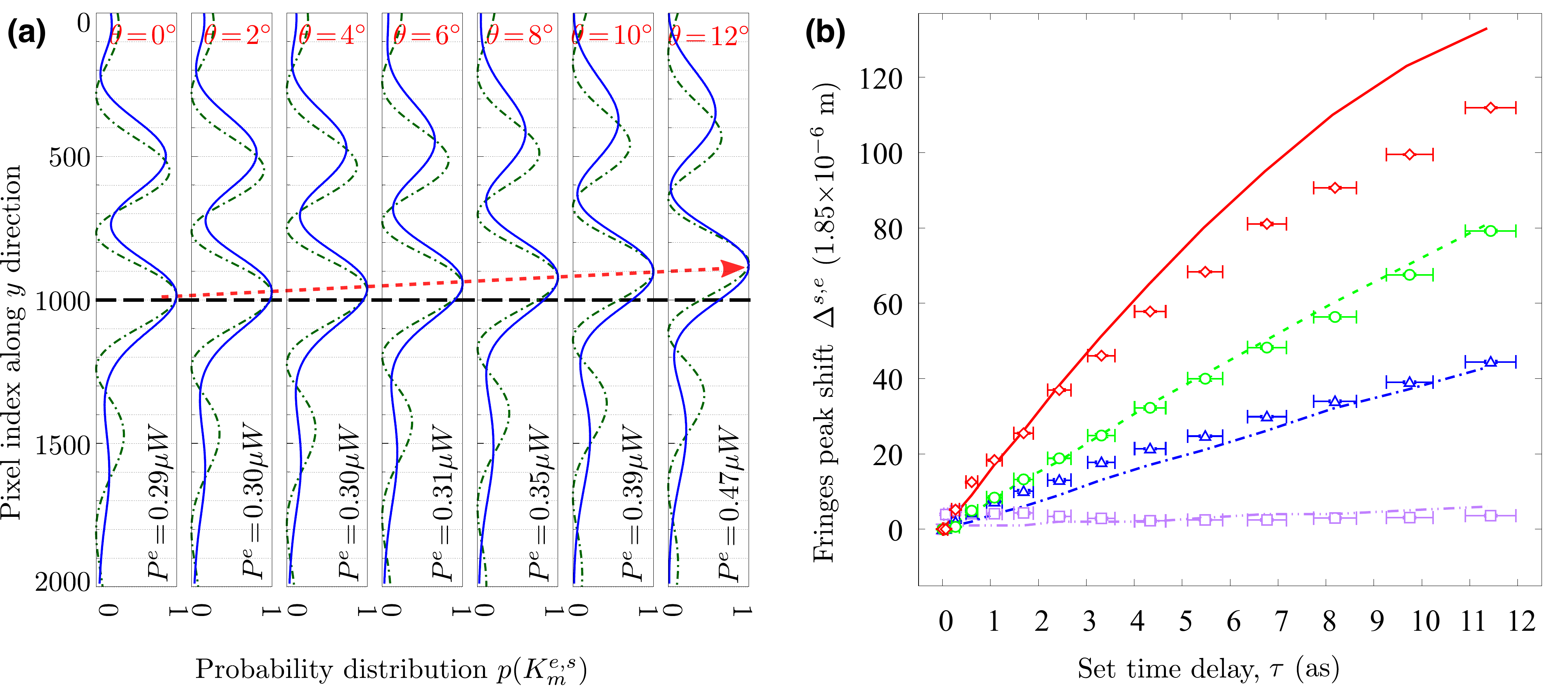}
\caption{\label{Fig:CCDiamgesOneDemention_1500as_inital_time_delay} 
Simulation and experimental results.
(a) The dependency of the probability distribution $p(K^{e,s}_{m})$ along $m-$direction pixels on $\theta$ when setting $\beta_{u,d}=\pm1.6^{\circ}$. 
The dashed (green) and solid (blue) lines correspond to the simulation and experimental results, respectively. The red arrow line is a guide for peak shifts. 
The black dotted line represents the horizontal line.
{
The simulation results contain an initial time delay $t_0$= 900 as in the double-slit interferometer.
(b) Fringe peak shifts $\Delta^{e,s}$ with varying time delay $\tau$ and various post-selection angle $\beta^{u,d}$.
Lines and points with error bars represent the theory and experimental results, respectively.
{The horizontal error bars correspond to the uncertainty of $\delta \theta= \pm 0.15^{\circ}$ in the set time delay limited by the precision of the Vernier scale used to adjust the waveplate angles; the vertical error bars correspond to the uncertainty in estimating the fringe shifts.}
The red, green, blue and purple data represent $\beta^{u,d}=\pm1.6^{\circ}$, $\beta^{u,d}=\pm3.3^{\circ}$, $\beta^{u,d}=\pm6.6^{\circ}$ and $\beta^{u,d}=\pm45^{\circ}$, respectively. 
}
}
\end{figure*}

We display both the expectations and the experimental probability distributions $p(K^{e,s}_{m} | \tau, X)$  along the $m-$direction with $\beta^{u,d}= \pm 1.6^{\circ}$ in Fig.~\ref{Fig:CCDiamgesOneDemention_1500as_inital_time_delay}(a). 
We use a sub-array of $2000\times405$ pixels in the $y$- and $x$- directions, respectively.
It is evident that the entire {interference fringes} shift with increasing time delay.
{The experimental curves vary with time delay qualitatively in the same way as the theoretical curves.
However, the experimental probability distribution $p(K_{m}^{e,s}$) presented in Fig.~\ref{Fig:CCDiamgesOneDemention_1500as_inital_time_delay}(a) is seen to have lower fringe visibility than is expected from the theoretical simulations. The main reason is that using two different post-selection angles will cause the two pointers to have different polarization states in the far field. In addition, aberrations caused by lenses in the 4-$f$ system and the focused lens could change the position and height of the fringes. 
We will further improve our experimental setup to reduce these imperfections in our future work.
}

Futhermore, Fig.~\ref{Fig:CCDiamgesOneDemention_1500as_inital_time_delay}(b) displays the experimentally observed fringe shift $\Delta^{e}$ alongside the theoretically simulated fringe shift $\Delta^{s}$.
The experimental results closely match the simulations when $\beta^{u,d}$ is set to $\pm3.3^{\circ}$ and $\pm6.6^{\circ}$. At the smallest postselection angle of $\beta^{u,d}=\pm1.6^{\circ}$, the experimental data are generally lower than the theoretical predictions.
The discrepancy in Fig.~\ref{Fig:CCDiamgesOneDemention_1500as_inital_time_delay}(b) between the red data points and curve at $\beta^{u,d}=\pm1.6^{\circ}$, is a phenomenon that will always arise in weak value amplification. To put it simply, the maximum amplification will, given a sufficiently small estimation parameter, be limited by experimental imperfections. These are imperfections in the initial state, post-selected state, and the interaction and they limit the orthogonality at the heart of large amplification. One ascertains the maximum possible amplification by increasing the amplification factor until discrepancies begin to impede the estimation of the target parameter. This approach is fairly standard in weak value amplification, for example see early work: Ben Dixon, David J. Starling, Andrew N. Jordan, and John C. Howell, “Ultrasensitive beam deflection measurement via interferometric weak value amplification,” Phys. Rev. Lett. 102, 173601 (2009)~\cite{PhysRevLett.102.173601}.

\section{Experimental signal-to-noise ratio}
{
The experimental precision in measuring time delay $\tau$ is quantitatively calculated by the signal-to-noise ratio (${\rm SNR}$) with statistical error.
For each estimation of the time delay, in order to obtain the SNR with the associated statistical error, the total measurements $N$ are divided into $k=N/n$ groups, where $n$ denotes the subgroup size.
Then, we calculate ${\rm SNR}_{k}$~\cite{PhysRevLett.126.220801} of each subgroup and ${\rm SNR}_{k}$ is quantitatively defined as
\begin{equation} \begin{split}
\label{Eq:Posson_distribution_simulation}
{\rm SNR}_{k}={\langle X_n \rangle}/{\sqrt{ {\rm Var} [X_n ]}} ,
\end{split} \end{equation}
where $X_n=(1/n) \sum_{i=1}^{n} \tau_{i}$ is the calibration result after $n$ times measurement. The signal mean value is $\langle X_n \rangle$ and ${\rm Var} [X_n ]$=${{\sigma}^{2}}$ represents the variance of the measured signal. 
Finally, we calculate the overall SNR with the mean value $\langle {\rm SNR}_{k} \rangle$ and standard deviation ${\sqrt{ {\rm Var} [{\rm SNR}_{k} ]}}$. 
}

\end{document}